\documentclass[prl,twocolumn,showpacs,amsmath,amssymb,superscriptaddress]
{revtex4}

\usepackage[dvips]{color}
\usepackage{graphicx}

\begin{document}

\title{Comment on ``Novel Superfluidity in a Trapped Gas of Fermi Atoms 
with Repulsive Interaction Loaded on an Optical Lattice''}

\author{M. Rigol}
\affiliation{Physics Department, University of California, Davis,
CA 95616, USA}
\author{S. R. Manmana}
\affiliation{Institut f\"ur Theoretische Physik III, Universit\"at 
Stuttgart, 70550 Stuttgart, Germany}
\affiliation{Fachbereich Physik, Philipps-Universit\"at Marburg, 
35032 Marburg, Germany}
\author{A. Muramatsu}
\affiliation{Institut f\"ur Theoretische Physik III, Universit\"at 
Stuttgart, 70550 Stuttgart, Germany}
\author{R. T. Scalettar}
\affiliation{Physics Department, University of California, Davis,
CA 95616, USA}
\author{R. R. P. Singh}
\affiliation{Physics Department, University of California, Davis,
CA 95616, USA}
\author{S. Wessel}
\affiliation{Institut f\"ur Theoretische Physik III, Universit\"at 
Stuttgart, 70550 Stuttgart, Germany}

\pacs{03.75.Ss, 05.30.Fk, 71.30.+h, 74.20.Mn}

\maketitle

In a recent letter \cite{machida04} (referred to as I below), 
Machida {\it et~al.} made the exciting claim that in a 
{\em one-dimensional} (1D) trapped gas 
of fermions with repulsive interactions a superfluid phase appears 
around the Mott-insulator (MI) at the center of the trap (COT). 
Their claim is based on a negative binding energy ($E_b$), 
and a large weight for a singlet formed by particles located at 
opposite sides of the MI. We show here that the observed effects 
are not related to superfluidity.

After a MI forms at large $U$, two particles with opposite spins added 
to the trap prefer to sit beyond the two ends of the MI phase in order 
to avoid double occupancy. Hence, the large weight of the singlet 
[Eq.\ (3)] in Fig.\ 4 of I can be understood to be a simple 
consequence of the density distribution and the antiferromagnetic 
character of the MI state, i.e., it does not signal superfluidity. 

We then focus on the origin of the negative $E_b$ observed in I. 
Most of the results in I exhibit a non-zero density at 
the borders of the trap, i.e, they depend on the boundary 
conditions. We thus recalculated two cases depicted in Fig.\ 1(a) of I, 
keeping the same curvature of the trap and increasing the system 
size to $N=20$. This ensures zero density at the borders, as it should 
be for confined systems. We used quantum Monte Carlo (QMC) simulations 
\cite{rigol03,rigol04}, density-matrix renormalization group (DMRG) 
\cite{dmrg}, and exact diagonalization (ED).

\begin{figure}[h]
\includegraphics[width=0.4\textwidth,height=0.26\textwidth]{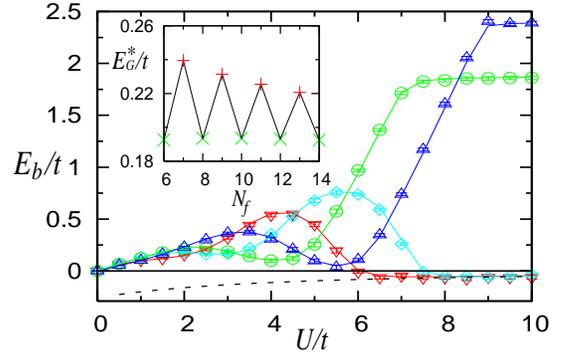}
\caption{(color online). 
QMC (points) and DMRG (continuous lines) results for
$E_b(10)=E(5,5)+E(6,6)-2E(5,6)$ vs $U/t$ for:
$V=7.5t$ (\textcolor{red}{$\bigtriangledown$},\textcolor{green}{$\bigcirc$}), 
and $V=9.5t$ (\textcolor{cyan}{$\Diamond$},\textcolor{blue}{$\triangle$}) 
(see I), when the center of the trap is located in the middle of two 
lattice points 
(\textcolor{red}{$\bigtriangledown$},\textcolor{cyan}{$\Diamond$}), 
and in a lattice point (\textcolor{green}{$\bigcirc$},\textcolor{blue}{$\triangle$}). 
The dashed line shows ED results for $E_b(10)$ in open MI systems 
(see text). The inset shows ED results for $E^*_G=E_G+AN_f$ vs $N_f$, 
in open MI systems with $U/t=8$. 
Here, we choose a nonzero  value of $A=0.327$ to stress the
even(\textcolor{green}{$\times$})-odd(\textcolor{red}{$+$}) 
effect, without affecting the actual value of 
$E_b<0$ it causes.}
\label{BE_Trap}
\end{figure}

Figure \ref{BE_Trap} shows that a negative $E_b$ appears for large 
values of $U/t$, at the point where the MI sets in the COT. 
Beyond this point adding more particles to the system causes the MI 
to increase in size. This is in contrast to the systems without a trap, 
where adding more particles to the half-filled case causes the MI to disappear. Both 
negative and positive $E_b$ arise in the latter doped case for different boundary conditions \cite{fye90}.

The dashed line in 
Fig.\ 1 corresponds to ED results of the 1D Hubbard model without trap, 
at half-filling, and open boundary conditions (OBC). 
Here $E_b$ is calculated by adding a site when adding a particle,
for OBC, in order to simulate the MI in the middle of the trap
without the metallic wings. The results obtained are 
practically indistinguishable from the $E_b$ obtained for 
trapped systems after the MI appears in the COT. Therefore,
the negative $E_b$ is due to the MI region, and does not signal superfluidity 
in the wings of the MI. Moreover, in the inset of Fig.\ \ref{BE_Trap}
we show that the negative $E_b$ in the MI is due to an even-odd effect. 
There we have plotted the ground-state energy $E_G$ for MI systems 
without the trap, and OBC, vs the number of particles ($N_f$). 
The even-odd effect is evident, and becomes smaller with increasing 
system size. Additionally, consistent with the results above, 
we find that: 
(i) Displacing the COT from the middle of two lattice points, 
as selected in I, leads to positive values of $E_b$. 
Results for the COT on a lattice point are also shown 
in Fig.\ \ref{BE_Trap}.
(ii) In the trap, similarly to the OBC case, the negative 
$E_b\rightarrow 0$ almost linearly with increasing system size, 
i.e, it is a finite size effect.

This work was supported by NSF-DMR-0312261, NSF-DMR-0240918, 
NSF-ITR-0313390, SFB 382, HLR-Stuttgart, and NIC at FZ J\"ulich.
We thank R.~M. Noack for helpful discussions.

\end{document}